\documentclass[pla,aps,twocolumn,ams]{revtex4} 
\usepackage{amsfonts}
\usepackage{amsmath}
\usepackage{amssymb}
\usepackage{graphicx}
\usepackage{dcolumn}
\usepackage{bm}
\usepackage{color}

\begin{document} 
 
\title{ 
Pair-breaking of multi-gap superconductivity under parallel magnetic fields in electric-field-induced surface metallic state
} 
\author{Masahiro Nabeta$^1$} 
\affiliation{
$^1$Department of Physics, Okayama University, Okayama 700-8530, Japan \\
$^2$Research Institute for Interdisciplinary Science (RIIS), Okayama University, Okayama 700-8530, Japan}
\author{Kenta K. Tanaka$^1$} 
\affiliation{
$^1$Department of Physics, Okayama University, Okayama 700-8530, Japan \\
$^2$Research Institute for Interdisciplinary Science (RIIS), Okayama University, Okayama 700-8530, Japan}
\author{Seiichiro Onari$^{1,2}$} 
\affiliation{
$^1$Department of Physics, Okayama University, Okayama 700-8530, Japan \\
$^2$Research Institute for Interdisciplinary Science (RIIS), Okayama University, Okayama 700-8530, Japan}
\author{Masanori  Ichioka$^{1,2}$} 
\affiliation{
$^1$Department of Physics, Okayama University, Okayama 700-8530, Japan \\
$^2$Research Institute for Interdisciplinary Science (RIIS), Okayama University, Okayama 700-8530, Japan}
\date{\today}

\begin{abstract}
Roles of paramagnetic and diamagnetic pair-breaking effects in 
superconductivity in electric-field-induced surface metallic state
are studied by Bogoliubov-de Gennes equation, 
when magnetic fields are applied parallel to the surface. 
The multi-gap states of sub-bands are related to the depth dependence and 
the magnetic field dependence of superconductivity. 
In the Fermi-energy density of states and the spin density, 
sub-band contributions successively appear from higher-level sub-bands 
with increasing magnetic fields. 
The characteristic magnetic field dependence may be a key feature 
to identify the multi-gap structure of 
the surface superconductivity.
\end{abstract}
 

\maketitle 
\section{Introduction} 

Electric-field-induced carrier-doping by 
the field-effect-transistor structure or 
the electric-double-layer-transistor (EDLT) 
structure~\cite{UenoJPSJ,Ueno,Ye,UenoKTO,Taniguchi,YeMoS2} 
is a new powerful method to control the carrier density 
by the gate voltage. 
At surfaces of insulators or semiconductors, 
carriers are induced near the surface by the strong electric field, 
and trapped in the confinement potential of the electric field. 
In the surface metallic states by the EDLT, 
superconductivity is realized 
at low temperatures~\cite{UenoJPSJ}, as performed in 
${\rm SrTiO_3}$~\cite{Ueno}, 
${\rm ZrNCl}$~\cite{Ye},  
${\rm KTaO_3}$~\cite{UenoKTO}, and 
${\rm MoS_2}$~\cite{Taniguchi,YeMoS2}.  
The surface superconductivity in ${\rm SrTiO_3}$ 
was also realized at the interface of 
${\rm LaTiO_3 / SrTiO_3}$
and ${\rm LaAlO_3 / SrTiO_3}$~\cite{Biscaras,Bert}. 

A unique nature of the surface metallic state is that 
sub-bands are formed by the quantum confinement of carriers near 
the surface~\cite{Ueno,Santander-Syro,King}.  
This is different nature from three-dimensional bulk metals or 
ideal two-dimensional systems. 
The surface superconductivity is expected to have 
multi-gaps depending on the sub-bands~\cite{Mizohata}. 
The sub-band dependent multi-gaps are tightly related to 
the spatial variation of the superconductivity along the depth direction. 
It is also suggested that crossover from the Bardeen-Cooper-Schrieffer (BCS) 
pair to the Bose-Einstein condensate (BEC) occurs 
in one of the sub-bands~\cite{Mizohata}.  
The multi-gap and the BCS-BEC crossover are universal features 
in superconductivity within nano-scale quantum 
confinement~\cite{Shanenko1,Shanenko2}. 
In many previous studies,  superconductivity 
in nano-scale quantum confinement was considered in the 
potential well $V({\bf r})=0$ 
inside of the confinement and $V({\bf r}) \rightarrow \infty$ 
at the boundary~\cite{Shanenko1,Shanenko2,Shanenko,Croitoru,Chen}.
However, in the surface superconductivity in the EDLT structure 
we have to consider the spatial variation of $V({\bf r})$ such as in 
the triangular potential. 
There wave functions of the electronic states were studied 
in the electric-field-induced surface metallic 
states~\cite{Ueno,Santander-Syro,King}.   
The superconductivity in this case is studied by 
a method of the Bogoliubov-de Gennes (BdG) equation, 
such as in Ref. ~\cite{BdG,Mizohata}.  
This method was also used in the theoretical 
studies~\cite{Mizushima,Takahashi} to explain the 
spatial structure of superfluidity of Fermionic atomic gas 
trapped in a harmonic potential~\cite{ZwierleinS,ZwierleinN}.

On the other hand, measurements of physical properties 
related to the pair-breaking by magnetic fields 
are important methods to know the character of each superconducting system. 
The surface superconductivity in the EDLT structure 
was identified as two-dimensional one, 
from the magnetic field orientation dependence of the upper critical 
field~\cite{UenoMag}. 
This indicates that vortices do not penetrate into 
narrow surface superconducting region, 
when magnetic field is applied parallel to the surface. 

The purpose of this paper is to study the pair-breaking 
of the surface superconductivity when magnetic field is 
applied parallel to the surface, 
and to clarify how influences of the 
sub-band dependent multi-gap superconductivity appear 
in the magnetic field $H$. 
We calculate properties of the surface superconductivity by   
the BdG equation~\cite{BdG,Mizohata}, 
assuming isotropic $s$-wave pairing. 
In addition to the paramagnetic pair-breaking by the Zeeman shift 
of the Fermi energy level between up and down spins, 
we study influences of the diamagnetic pair-breaking by 
the screening current to applied magnetic fields. 
In Ref. \cite{Nabeta}, only the paramagnetic pair-breaking effect was 
studied in the case of confinement potential 
including screening effect of electric fields by carriers. 
Since diamagnetic pair-breaking effect was not considered 
in previous studies, 
it is necessary to clarify contributions of diamagnetic effects 
as another mechanism of pair-breaking under parallel magnetic fields.

This paper is organized as follows. After the introduction, 
we explain our theoretical formulation of the BdG equation 
under parallel magnetic fields in Sec. \ref{sec:formulation} 
and Appendix \ref{sec:formulation2}. 
We study influences of magnetic fields to the pair potential in Sec. \ref{sec:PairPotential}. 
Section \ref{sec:Current} is for current, spin current, and internal field. 
As phenomena reflecting multi-gap superconductivity of sub-band system, we study the magnetic field dependence of electronic states in the surface superconductivity in Sec. \ref{sec:ElectronicState}, and paramagnetic spin density in Sec. \ref{sec:SpinDensity}. 
The last section is devoted to the summary.

\section{Bogoliubov-de Gennes equation under parallel magnetic fields} 
\label{sec:formulation}

In this paper, 
we calculate the pair potential $\Delta({\bf r})$, 
and the wave functions  
$u_{\epsilon}({\bf r})$, $v_{\epsilon}({\bf r})$
for the eigen-energy $E_\epsilon$ 
by the BdG equation~\cite{BdG}. 
Notes on the derivation of the BdG equation and the related equations 
in the presence of paramagnetic effect are described in 
Appendix \ref{sec:formulation2}. 
In our coordinate ${\bf r}=(x,y,z)$, 
$z$-axis is the depth direction perpendicular to the surface at $z=0$. 
As the confinement potential near the surface, 
we use the triangular potential, 
$V(z)=|e|F_0 z$ for $z>0$ and $V(z) \rightarrow \infty$ for $z \le 0$.
We typically consider the case of  
sheet carrier density 
$n_{\rm 2D}=6.5 \times 10^{13}$[${\rm cm^{-2}}$],
electric field $F_0 =1.4 \times 10^{-3}$[${\rm V/nm}$] at the surface, and  
single band with effective mass $m^\ast=4.8m_0$,   
where $m_0$ is free electron's mass. 
This corresponds to 
one of the cases studied in Ref. \cite{Mizohata} considering ${\rm SrTiO_3}$.

We set the vector potential as 
${\bf A}=(0,A_y,0)$ with $A_y=-Hz$, so that the magnetic field 
parallel to the surface is applied along the $x$ direction, 
and the screening current flows along the $y$ direction. 
In this situation, we can set 
$\Delta({\bf r})=\Delta(z){\rm e}^{{\rm i}2qy}$ and 
\begin{eqnarray}
\left(\begin{array}{c}
u_{\epsilon}({\bf r}) \\ 
v_{\epsilon}({\bf r}) \\ 
\end{array}\right) 
=
\frac{1}{\sqrt{S}}{\rm e}^{{\rm i}(k_x x + k_y y) }
\left(\begin{array}{c}
u_{\epsilon}(z) {\rm e}^{ {\rm i}qy} \\ 
v_{\epsilon}(z) {\rm e}^{-{\rm i}qy} \\ 
\end{array}\right),   
\label{eq:uv-funs}
\end{eqnarray} 
where $S$ is unit area of surface. 
$q$ is a constant relating to the current flow, 
where we assume that physical quantities do not have $y$-dependence. 
We do not consider the penetration of vortices. 
We decide $q$ value from the current conservation, 
as explained later in this section.   
Thus the BdG equation is reduced to 
\begin{eqnarray} 
\left(\begin{array}{cc}
K_+ & \Delta(z) \\ 
\Delta(z) & -K_- \\ 
\end{array} \right) 
\left(\begin{array}{c}
u_{\epsilon}(z) \\ 
v_{\epsilon}(z) \\ 
\end{array}\right) 
=E_{\epsilon} 
\left(\begin{array}{c}
u_{\epsilon}(z) \\ 
v_{\epsilon}(z) \\ 
\end{array}\right) ,  
\label{eq:BdG1}
\end{eqnarray} 
with the kinetic term 
\begin{eqnarray} && 
K_\pm 
=\frac{\hbar^2}{2m^\ast}\left(k_x^2+(\pm k_y +q +\frac{\pi}{\phi_0}A_y)^2 
-\partial_z^2 \right)
\nonumber \\ && \qquad 
+V(z) \mp \mu_{\rm B}H -\mu ,
\end{eqnarray} 
where $\phi_0$ is a flux quantum, and $\pm\mu_{\rm B}H$ is Zeeman energy 
with the Bohr magneton $\mu_{\rm B}=5.7883 \times 10^{-5}$ [eV/T].  
The chemical potential $\mu$ ($\sim E_{\rm F}$: Fermi energy) 
is determined to fix $n_{\rm 2D}$. 
The eigen-states of Eq. (\ref{eq:BdG1}) are 
labeled by 
$\epsilon \equiv (k_x,k_y,i_z,\alpha)$. 
$i_z\ (=1,\ 2, \ \cdots)$ indicates 
label for sub-bands coming from quantization by 
confinement in the $z$-direction. 
$\alpha$ is for two states of particle and hole branches.
As the boundary condition at the surface, 
we set $u_{\epsilon}(z)=v_{\epsilon}(z)=0$. 
In the following, energy, length, magnetic field, and 
local carrier densities are, 
respectively, presented in unit of ${\rm meV}$, ${\rm nm}$, ${\rm T}$, 
and ${\rm nm}^{-3}$. 

The pair potential is calculated by the gap equation
\begin{eqnarray} && 
\Delta(z)
=V_{\rm pair} {\sum_{\epsilon}}'  u_{\epsilon}(z) v_{\epsilon}(z) f(-E_{\epsilon}) 
\label{eq:Delta1}
\end{eqnarray}
with the Fermi distribution function $f(E)$.  
In Eq. (\ref{eq:Delta1}), the energy cutoff $ E_{\rm cut} $
of the pairing interaction is considered in the summation 
as ${\sum_\epsilon}'=\sum_\epsilon \theta(E_{\rm cut}-|E_\epsilon|$) 
using the step function $\theta$.
Here, we consider a conventional case 
of spin-independent isotropic $s$-wave pairing. 
We typically use $V_{\rm pair}=0.08$, and $E_{\rm cut}=$10 meV. 
These give the transition temperature $T_{\rm c} \sim 6.3 {\rm K}$. 
We consider this larger $T_{\rm c}$ case to ensure 
energy resolution within the superconducting gap in our calculations. 
Therefore, gap amplitude and critical field are about 15 times larger 
than those in superconductivity of ${\rm SrTiO_3}$~\cite{Ueno,UenoMag}, 
but qualitative behaviors are not significantly changed. 
While the coherence length is expected to become $\sqrt{15}$ times shorter, 
the $z$-dependence of $\Delta(z)$ is determined  by 
the spatial variation of the wave functions in the confinement potential, 
rather than the coherence length, as shown later. 
Thickness of surface superconducting region is still narrow, 
so that we neglect penetration of vortices with core radius 
in the order of the coherence length.

The local carrier density $n(z)=n_\uparrow(z)+n_\downarrow(z)$, 
spin density $m(z)=n_\uparrow(z)-n_\downarrow(z)$, 
current density $J(z)=J_\uparrow(z)+J_\downarrow(z)$, 
and spin current density $J_s(z)=J_\uparrow(z)-J_\downarrow(z)$ 
are calculated from up- and down-spin contributions 
\begin{eqnarray} && 
n_{\uparrow}(z)=\sum_{\epsilon} |u_{\epsilon}(z)|^2 f(E_{\epsilon}), 
\\ && 
n_{\downarrow}(z)=\sum_{\epsilon} |v_{\epsilon}(z)|^2 f(-E_{\epsilon}), 
\\ && 
J_{\uparrow}(z)
=\frac{e \hbar}{m^\ast}
\sum_{\epsilon} \left(k_y+q+\frac{\pi}{\phi_0}A_y\right)
                |u_{\epsilon}(z)|^2 f(E_{\epsilon}), 
\quad 
\label{eq:J-up}
\\ && 
J_{\downarrow}(z)
=\frac{e \hbar}{m^\ast}
\sum_{\epsilon} \left(-k_y+q+\frac{\pi}{\phi_0}A_y\right)
                |v_{\epsilon}(z)|^2 f(-E_{\epsilon}).
\quad 
\label{nz}
\end{eqnarray} 
Derivations of Eqs. (\ref{eq:Delta1})-(\ref{nz}) 
are explained in Appendix \ref{sec:formulation2}. 

Density of states (DOS) with the spin-decomposition is obtained as 
\begin{eqnarray}
N(E)=N_\uparrow(E)+N_\downarrow(E)=\int_0^\infty N(E,z){\rm d}z
\end{eqnarray}
from the local DOS $N(E,z)=N_\uparrow(E,z)+N_\downarrow(E,z)$ by 
\begin{eqnarray} && 
N_{\uparrow}(E,z)
=\sum_{\epsilon} |u_{\epsilon}(z)|^2 \delta(E-E_{\epsilon}) ,  
\\ && 
N_{\downarrow}(E,z)
=\sum_{\epsilon} |v_{\epsilon}(z)|^2 \delta(E+E_{\epsilon}) . 
\end{eqnarray} 

To identify roles of paramagnetic and diamagnetic pair-breaking effects, 
calculations are performed in two cases. \\
Case (i): Only the paramagnetic pair-breaking is 
considered by setting $A_y=q=0$. \\
Case (ii): Both diamagnetic and paramagnetic pair-breakings are 
considered. 
We set $A_y=-Hz$. 
From Eqs. (\ref{eq:J-up}) and (\ref{nz}), 
total current $J_{\rm total}\equiv \int_0^\infty J(z){\rm d}z$   
is an increasing function of $q$. 
We decide $q$ value so that it satisfies the current conservation 
$J_{\rm total}=0$.  

In our numerical calculations, we discretize the region $0 <z< 30$ 
to $N_z=151$ points, 
and $\partial_z$ is estimated by differences between neighbor points. 
Thus we calculate eigen-states of $2N_z \times 2N_z$ matrix 
in Eq. (\ref{eq:BdG1}) under given $(k_x,k_y)$. 
Since components of matrix in  Eq. (\ref{eq:BdG1}) are real, 
wave functions $u_{\epsilon}(z)$ and $v_{\epsilon}(z)$ are real functions. 
Iterating calculations of Eqs. (\ref{eq:BdG1}) and (\ref{eq:Delta1}) 
from an initial state with constant $\Delta(z)=$1.0 (i) or  0.82 (ii),  
we obtain selfconsistent results of $\Delta(z)$ and wave functions. 
We study the $H$-dependence of the superconducting state at 
a low temperature $T=1.16 \times 10^{-2}{\rm K} \ll T_{\rm c}$. 

\section{Pair potential} 
\label{sec:PairPotential}

\begin{figure}[tb] 
\begin{center}

\includegraphics[width=6.0cm]{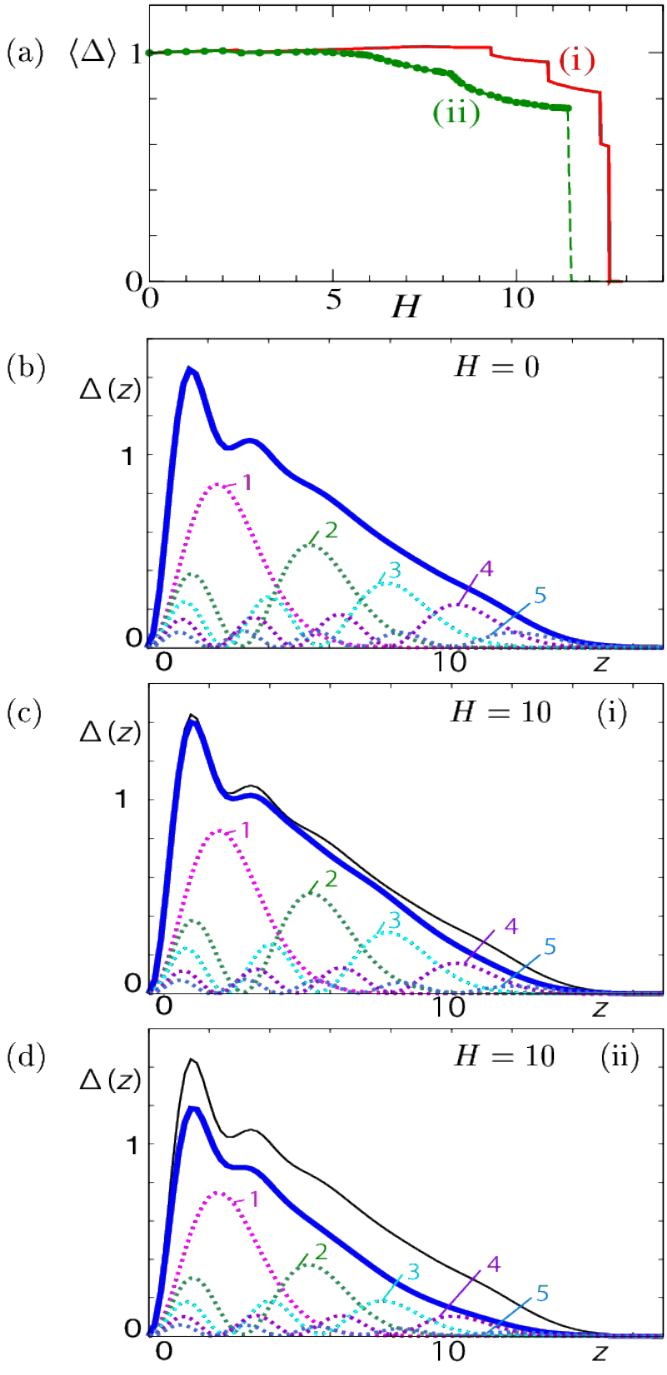} 

\end{center}
\caption{ 
(a) $H$-dependence of the average $\langle \Delta \rangle$ 
in the cases (i) and (ii).  
Dashed line of (ii) indicates possible first order transition of $H_{\rm c}$.
(b) Depth $z$-dependence of the pair potential $\Delta(z)$ (bold solid lines)
and the sub-band decompositions (dashed lines) to $i_z=1, \cdots, 5$. 
$H=0$. 
(c) The same as (b), but $H=10$ in the case (i) of only paramagnetic pair-breaking effect. 
(d) The same as (b), but $H=10$ in the case (ii) of both paramagnetic and diamagnetic pair-breaking effects. 
In (c) and (d), thin lines are for $H=0$ for comparison.
} 
\label{fig1} 
\end{figure} 

First, we study influences of $H$ on $\Delta(z)$. 
In Fig. \ref{fig1}(a) we plot the average 
$\langle \Delta \rangle \equiv 
\int_0^\infty \Delta(z) n(z) {\rm d}z /\int_0^\infty n(z) {\rm d}z$ 
as a function of $H$. 
At lower fields, 
$\langle \Delta \rangle$ is almost constant. 
In the case (i) of only paramagnetic effect, 
$\langle \Delta \rangle$ shows step-like suppression at $H > 9$. 
It suddenly vanishes at the first order transition of the critical field $H_{\rm c} \sim 12.5$. 
These are similar behavior to those suggested in superconductivity 
within nano-scale quantum confinement~\cite{Shanenko,Croitoru,Chen}. 
In our calculation, the possibility of FFLO~\cite{FF,LO} states is not considered. 
In the case (ii), the step-like change is smeared, and further suppression occurs by the additional diamagnetic pair breaking at $H>5$.  
This indicates that diamagnetic pair breaking effect can be 
another important mechanism for suppression of superconductivity 
even under parallel magnetic fields. 
In Fig. \ref{fig1}(a),  we present dashed line 
to indicate possible first order transition of $H_{\rm c}$ 
in the case (ii), 
since $\Delta(z) \rightarrow 0$ at $H \ge 11.5$ 
in the calculation starting from an initial state with small $\Delta$. 

The depth dependences of $\Delta(z)$ and the sub-band decomposition are presented in Figs. \ref{fig1}(b)-\ref{fig1}(d). 
Reflecting the confinement in $V(z)$, 
$\Delta(z)$ has a peak at $z \sim 1.2$, and decreases towards zero at $z \sim 15$. 
With increasing sub-band index $i_z$, amplitudes of contributions to $\Delta(z)$ become smaller, but extend to deeper $z$ region.   
Compared with the case of a zero-field in Fig. \ref{fig1}(b), 
$\Delta(z)$ is weakened at deeper positions for $H=10$ in Fig. \ref{fig1}(c), 
because suppression of superconductivity occurs
only in higher-level sub-bands $i_z=4$ and 5, 
whose contributions extend to deeper region. 
In the case (ii) in Fig. \ref{fig1}(d), since the suppression occurs 
at sub-bands $i_z=2,\cdots, 5$, $\Delta(z)$ shows suppression 
at all $z$ region. 
Therefore, sub-band contributions are related to the depth dependence of the superconducting state.

\section{Current and spin current} 
\label{sec:Current}

\begin{figure}[tb] 
\begin{center}

\includegraphics[width=6.0cm]{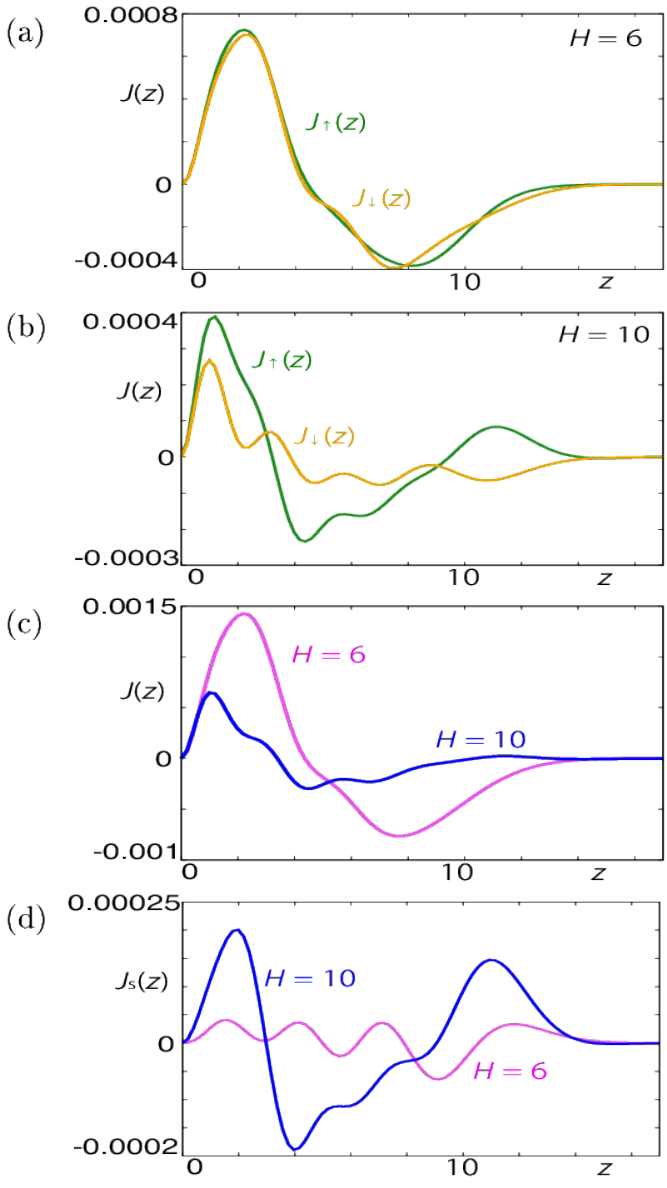} 

\end{center}
 \caption{ 
Depth $z$ dependence of current. 
(a) Spin-dependent current $J_\uparrow(z)$ and $J_\downarrow(z)$ at $H=6$.  
(a) $J_\uparrow(z)$ and $J_\downarrow(z)$ at $H=10$.  
(c) Total current $J(z)$ at $H=6$ and 10. 
(d) Spin current $J_{\rm s}(z)$ at $H=6$ and 10. 
Vertical axis is in unit of $e \hbar/{m^\ast}$. 
} 
\label{fig2} 
\end{figure} 

In the case (ii) including diamagnetic pair-breaking effect, 
screening current flows along the $y$ direction parallel to the surface. 
In this section, we study the current, spin current, internal field, and diamagnetic magnetization. 
Figure \ref{fig2}(a) shows depth dependence of spin-dependent current $J_\uparrow(z)$ and $J_\downarrow(z)$ at $H=6$. 
There, we see small deviations between $J_\uparrow(z)$ and $J_\downarrow(z)$. 
At a high field $H=10$ near $H_{\rm c}$, the difference between 
$J_\uparrow(z)$ and $J_\downarrow(z)$ becomes larger, 
as shown in Fig. \ref{fig2}(b). 
The depth dependence of total current $J(z)=J_\uparrow(z)+J_\downarrow(z)$ is presented in Fig. \ref{fig2}(c) at $H=6$ and 10. 
The current $J(z)$ flows in order to 
screen penetration of magnetic fields from outside of thin superconducting region. 
Therefore, sign of $J(z)$ changes between near the surface and deeper $z$ region. 
The amplitude of $J(z)$ becomes weaker at a high field $H=10$, as the superconducting pair potential is suppressed by the parallel magnetic field. 
Spin current $J_s(z)=J_\uparrow(z)-J_\downarrow(z)$ appears in the presence of both paramagnetic and diamagnetic pair-breaking effects.  
The depth dependence of the spin current $J_s(z)$ is presented in Fig. \ref{fig2}(d) at $H=6$ and 10. 
Amplitude of $J_s(z) $ becomes larger, with increasing $H$. 
Imbalance of up- and down-spins occurs where superconductivity is partially suppressed. 
Therefore, the oscillating behavior of $J_s(z)$ at $H=6$ reflects oscillation of wave function in the higher-level sub-band. 

\begin{figure}[tb] 
\begin{center}

\includegraphics[width=6.0cm]{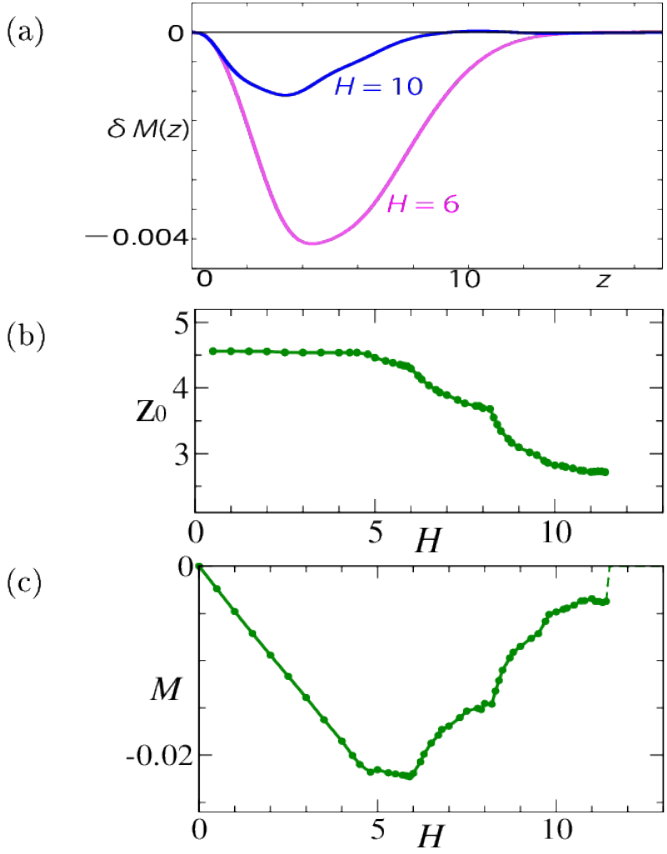} 

\end{center}
 \caption{ 
(a) Variation of internal field.  
Depth $z$-dependence of $\delta M(z)$ is presented at $H=6$ and 10. 
(b) $H$-dependence of $z_0 \equiv q \phi_0 / \pi H$ to satisfy 
the current conservation. 
(c)
$H$-dependence of magnetization $M$. 
Vertical axis is in unit of $e \hbar/ m^\ast$. 
} 
\label{fig3} 
\end{figure} 

From the total current $J(z)$ in Fig. \ref{fig2}(c), 
we calculate the variation of internal field as 
$\delta M(z) =- \int_0^z J(z'){\rm d}z'$. 
The depth dependence of $\delta M(z)$ is presented in Fig. \ref{fig3}(a). 
There, the internal field is slightly suppressed 
inside the superconducting region. 
At a high field $H=10$, 
width of superconducting region becomes smaller than that at $H=6$. 
Using a definition $z_0 \equiv q \phi_0 / \pi H$, we can write as 
$q + \frac{\pi}{\phi_0}A_y = -\frac{\pi}{\phi_0}H(z-z_0)$. 
For $q$ determined by the condition $J_{\rm total}=0$, 
we plot $z_0$ as a function of $H$ in Fig. \ref{fig3}(b). 
Since $z_0$ is located near the center of superconducting region, 
$z_0$ becomes smaller at higher $H$. 

The $H$-dependence of diamagnetic magnetization $M=\int_0^\infty \delta M(z) {\rm d}z$ is shown in Fig. \ref{fig3}(c). 
There, $M$ decreases linearly at low fields. 
And it increases towards zero at high fields, reflecting suppression of the pair potential by parallel magnetic fields. 
We note that $M$ is tiny quantity, 
since the superconducting region is narrow 
compared to the penetration length in the surface superconductivity.

\section{Magnetic field dependence of  electronic states} 
\label{sec:ElectronicState}

\begin{figure}[tb] 

\includegraphics[width=6.0cm]{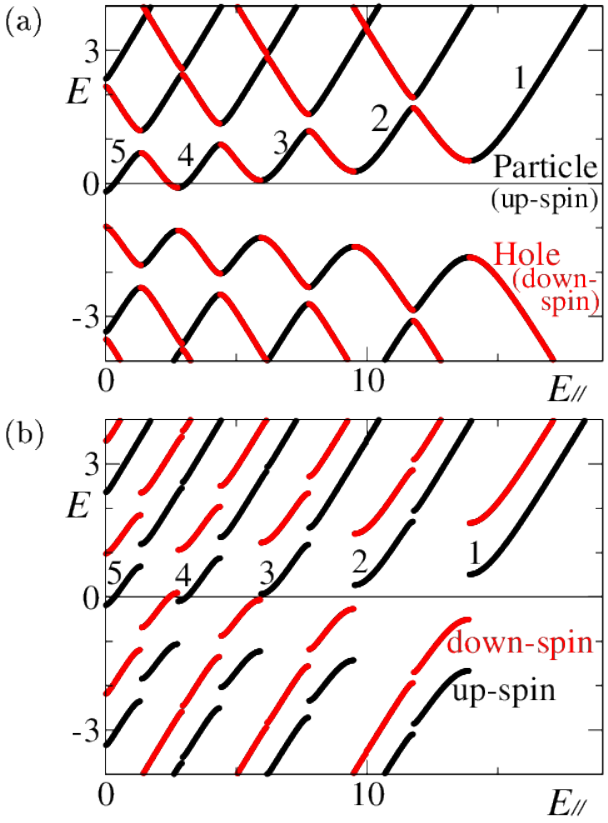} 

\caption{ 
(a) Eigen energy $E_\epsilon$ as a function of $E_\parallel$ 
for particle (up-spin) and hole (down-spin) branches of 
sub-bands $i_z=1, 2, \cdots$ at $H=10$ in the case (i). 
Fermi energy $E_{\rm F}$ corresponds to $E=0$. 
(b) 
Hole branches of down-spin are converted to the particle branches
in order to show dispersion of up- and down-spin electrons 
for sub-bands $i_z$. 
} 
\label{fig4} 
\end{figure} 

As phenomena where contributions of multi-gap superconductivity in the sub-band system clearly appear, 
we study influences of magnetic fields in the electronic states. 
In Fig. \ref{fig4}(a), we show eigen-energies $E_\epsilon$ as a function of 
$E_\parallel=\hbar^2 (k_x^2+k_y^2) / 2m^\ast $ 
in the case (i). 
For each sub-band $i_z=1,2,\cdots$, 
there exist two states of particle and hole branches. 
Line segments with positive (negative) slope are particle (hole) branches 
with $\int_0^\infty(|u_\epsilon(z)|^2 -|v_\epsilon(z)|^2){\rm d}z >0$ ($<0$), 
where main contributions come from up-spin electron's $K_+$ 
(down-spin hole's $-K_-$). 
At the energy where particle and hole branches cross each other 
in the normal state as $K_+=-K_-$, 
superconducting gap opens at each sub-band. 
The gap amplitude becomes smaller in higher-level sub-band, 
indicating multi-gap superconductivity~\cite{Mizohata}. 
The superconductivity in the sub-band $i_z=5$ is BEC-like~\cite{Mizohata,Shanenko1,Shanenko2}, since the gap is located at the bottom of the dispersion curve. 
The center energy of the gap moves from $E=0$ because of the Zeeman energy. 
Small gaps also appear at the crossing points of electron and hole 
branches between different sub-bands at higher $|E|$. 

In Fig. \ref{fig4}(b), hole branches of down-spin electrons are converted to 
the particle branch as $E_\epsilon \rightarrow -E_\epsilon$. 
This figure clearly shows the contribution of the Zeeman shift, i.e., 
the branches for up- (down-) spin shift to lower (higher) energy.  
In Fig. \ref{fig4} (b) at a high field $H=10$, 
the Fermi energy $E_{\rm F}$ 
is within the superconducting gap of lower-level sub-bands $i_z$=1 and 2. 
However, $E_{\rm F}$ is outside of gap-edge 
in the higher-level sub-band $i_z$=4, 
because the small gap in the sub-band is smaller than the Zeeman shift energy. 
When the Zeeman shift energy becomes the same order to 
the largest gap of the lowest sub-band $i_z=1$, 
the superconductivity vanishes at $H_{\rm c}$. 

\begin{figure}[tb] 
\begin{center}

\includegraphics[width=8.5cm]{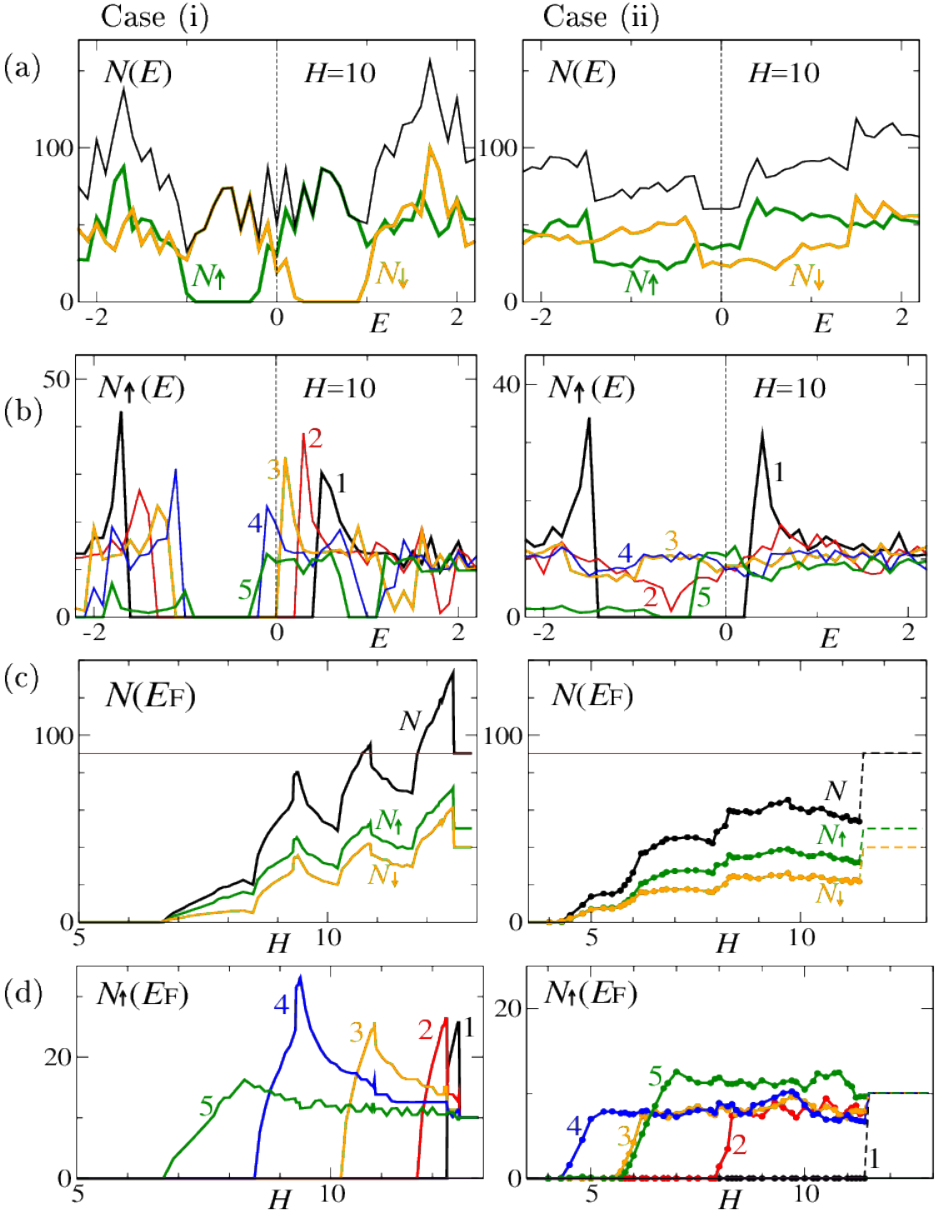} 

\end{center}
\caption{ 
Electronic state at $H=10$. 
Left-side and right-side panels are, respectively, for the cases (i) and (ii). 
(a) DOS $N(E)$ with the spin-resolved DOS $N_\uparrow(E)$ and $N_\downarrow(E)$. 
(b) Sub-band decompositions ($i_z=1$, $\dots$, 5) of $N_\uparrow(E)$.
(c) $H$-dependence of Fermi-energy DOS  
$N(E_{\rm F})$ with the spin-decompositions $N_\uparrow(E_{\rm F})$ 
and $N_\downarrow(E_{\rm F})$. 
$N(E_{\rm F})$ in the normal state is presented by a straight line. 
(d) Sub-band decompositions of $N_\uparrow(E_{\rm F})$ in (c). 
} 
\label{fig5} 
\end{figure} 

These behaviors by the Zeeman shift are reflected also 
in the DOS $N(E)$ in Fig. \ref{fig5}(a). 
There, finite DOS appears at $E_{\rm F}$¡¡
by the Zeeman shift of $N_\uparrow(E)$ and $N_\downarrow(E)$.  
We show the sub-band decompositions in Fig. \ref{fig5}(b). 
In the case (i), 
each sub-band has finite gap with sharp peak at the gap-edge. 
Higher-level sub-bands have smaller gap, 
and the gap-edges touch $E_{\rm F}$ by the Zeeman shift. 
Therefore Fermi-energy DOS $N(E_{\rm F})$ comes from higher-level sub-bands. 
On the other hand, in the case (ii), 
the gap is smeared due to the diamagnetic pair-breaking effect, 
and sharp peaks at the gap-edge vanish, except for $i_z=1$. 
Thus, $N_\uparrow(E)$ and $N_\downarrow(E)$ are gapless in Fig. \ref{fig5}(a). 

In Fig. \ref{fig5}(c), 
we plot $N(E_{\rm F})$ with $N_\uparrow(E_{\rm F})$ and $N_\downarrow(E_{\rm F})$   
as a function of $H$.
Sub-band decompositions of $N_\uparrow(E_{\rm F})$ are presented 
in Fig. \ref{fig5}(d).
We note that contribution of lowest sub-band $i_z=1$ 
does not appear until near $H_{\rm c}$.  
In the case (i) of Fig. \ref{fig5}(c), 
$N(E_{\rm F})$ appears at $H > 6$, 
and increases with multiple sharp peak behavior as a function of $H$. 
These peaks in the case of triangular confinement potential 
are contrasted to the behavior in the case of Ref. \cite{Nabeta}. 
The peak in Fig. \ref{fig5}(c) occurs 
when Zeeman energy $\mu_{\rm B} H$ equals the gap amplitude of a sub-band, 
and $E_{\rm F}$ touches sharp peak at the gap-edge in Fig. \ref{fig5}(b). 
Therefore, with increasing $H$, new contributions of lower-level sub-bands 
appear near 
the peak fields, as seen by 
the sub-band decomposition in Fig. \ref{fig5}(d). 
The multiple peak behavior of $N(E_{\rm F})$ in Fig. \ref{fig5}(c) 
is also a reason for the step of $\langle \Delta \rangle$ 
in the $H$-dependence in Fig. \ref{fig1}(a) for the case (i).  
For the case (ii) in Fig. \ref{fig5}(c), 
peak-behavior is smeared in the $H$-dependence of $N(E_{\rm F})$, 
because the gap-edges in $N(E)$ are smeared in Fig. \ref{fig5}(b) 
by the screening current at high fields.   
Thus, sub-band resolved DOS in Fig. \ref{fig5}(d) is not largely enhanced 
when it appears on raising $H$ in the case (ii). 

If observations about the $H$-dependence of DOS in Fig. \ref{fig5} are realized, 
such as by point contact tunneling junction at the surface, we may examine the multi-gap structure of sub-bands in 
the superconductivity of electric-field-induced surface metallic state, 
from the detailed structure of the $H$-dependence.
The differences between cases (i) and (ii) may be used 
to estimate ratio of the diamagnetic pair-breaking effect to 
the paramagnetic one from the experimental data. 
The ratio may be changed by material parameter and experimental conditions. 
When effective mass $m^\ast$ is larger as in the present calculations, 
or when superconducting region near surface is thinner by tuning gate voltage, 
diamagnetic pair-breaking effect is weakened, 
and paramagnetic pair-breaking effect becomes dominant.
In opposite case of smaller $m_0$ or thicker superconducting region, 
diamagnetic pair-breaking effect becomes important, 
where the first order transition at $H_{\rm c}$ may change to 
the second order transition.

\section{paramagnetic spin density}
\label{sec:SpinDensity}

\begin{figure}[tb] 
\begin{center}

\includegraphics[width=6.0cm]{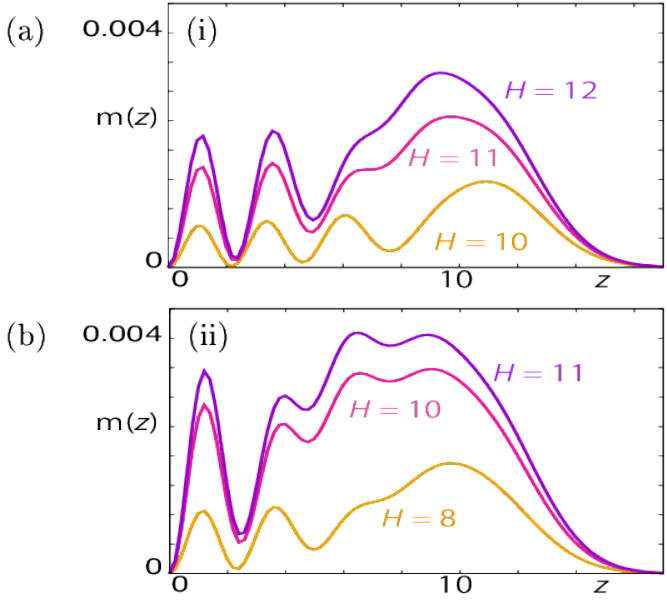} 

\end{center}
 \caption{ 
(a) Depth $z$ dependence of local spin density $m(z)=n_\uparrow(z)-n_\downarrow(z)$ at $H=10$, 11 and 12 in the case (i). 
(b) The same as (a) but for $H=8$, 10 and 11 in the case (ii). 
} 
\label{fig6} 
\end{figure} 

The sub-band contributions are also seen in paramagnetic spin density $m(z)$.   
Due to the Zeeman shift of the paramagnetic effect, there appears finite $m(z)$, 
corresponding to the Knight shift.   
Figure \ref{fig6}(a) presents the spatial variation of $m(z)$ in the case (i) for some fields $H$. 
Since imbalance of up- and down-spins comes from the suppressed region of the sub-band dependent superconductivity, $m(z)$ is larger at deep region by the higher-level sub-band contributions. 
The oscillating behavior of $m(z)$ at $H=10$ reflects 
oscillation of wave functions of higher-level sub-bands $i_z=4$ and 5. 
With increasing $H$, contributions of lower-level sub-bands also appear in addition to those of higher-level sub-bands, so that oscillating behavior of $m(z)$ is smeared at deep region. 
Figure \ref{fig6}(b) shows $m(z)$ in the case (ii). 
There, we see similar oscillating behaviors as in Fig. \ref{fig6}(a). 
In the case (ii),  $m(z)$ is larger from low fields, 
since the pair potential $\langle \Delta \rangle$ is smaller 
by the diamagnetic pair-breaking in addition to the paramagnetic pair-breaking. 

\begin{figure}[tb] 
\begin{center}

\includegraphics[width=6.0cm]{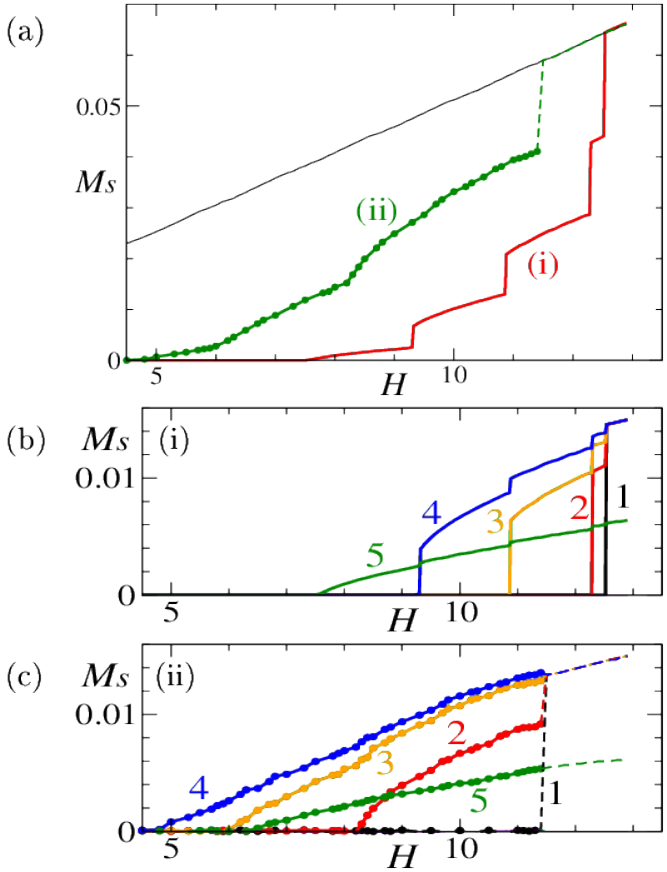} 

\end{center}
\caption{ 
(a) $H$-dependence of total paramagnetic moment $M_s$ 
in the cases (i) and (ii). 
A straight line is for $M_s$ in the normal state. 
(b) Sub-band decompositions ($i_z=1$, $\dots$, 5) of $M_s$ in the case (i) 
as a function of $H$. 
(c) The same as (b) but for the case (ii).   
} 
\label{fig7} 
\end{figure} 

Next, we present 
$H$-dependences of total paramagnetic moment $M_s=\int_0^\infty m(z) {\rm d}z$ 
in Fig. \ref{fig7}(a), 
and the sub-band decompositions in Figs. \ref{fig7}(b) and \ref{fig7}(c).
$M_s$ is zero at low fields, and appears from the middle fields. 
There, $M_s$ is constructed of contributions from higher-level sub-bands. 
In the case (i), the step-like increase of $M_s$ occurs 
when contributions of lower-level sub-bands are added as seen in Fig. \ref{fig7}(b). 
Also in the case (ii), contributions of lower-level sub-bands are 
successively added with increasing $H$ as presented in Fig. \ref{fig7}(c). 
However, step-like increase is smeared. 
This behavior comes from the smearing of gap structure in Fig. \ref{fig5}(b) by the diamagnetic pair-breaking due to the screening current. 

\section{Summary} 
\label{sec:Summary}

Roles of paramagnetic and diamagnetic pair-breaking effects by 
parallel magnetic fields were evaluated in superconductivity of 
the electric-field-induced surface metallic state, 
based on calculations of the BdG equation. 
There the depth-dependences of pair potential, current, spin current, internal field, and paramagnetic spin density  
were understood by the sub-band contributions. 
With increasing magnetic fields $H$, 
electronic states of sub-bands become normal-state-like 
successively from higher-level sub-bands to lower-level ones, reflecting multi-gap superconductivity. 
This is reflected in the $H$-dependence of Fermi-energy DOS $N(E_{\rm F})$ and 
total paramagnetic moment $M_s$. 
We found that 
steps or peaks in the $H$-dependence due to the paramagnetic pair-breaking 
are smeared by the diamagnetic pair breaking effect, 
because superconducting gap in higher-level sub-bands becomes gapless by the contributions of the screening current.   
As only paramagnetic pair-breaking effect was studied without including 
diamagneitc effect in previous theoretical studies, 
the present study showed that 
diamagnetic screening current induces another important pair-breaking effect 
even under parallel magnetic fields.
We expect that observation of the $H$-dependence, 
such as by point contact tunneling junction at the surface,  may be a clue to 
examine the multi-gap structure and the pair-breaking effects 
by magnetic fields in superconductivity of 
the electric-field-induced surface metallic state.

\section*{Acknowledgement}

This work was supported by JSPS KAKENHI Grant Number 25400373. 

\appendix

\section{
BdG equation in the presence of  paramagnetic effect 
}
\label{sec:formulation2}

If we consider all possible spin pairings, BCS Hamiltonian is given by $4 \times 4$ matrix 
with a base $(\hat\psi_\uparrow, \hat\psi_\downarrow, \hat\psi^\dagger_\uparrow, \hat\psi^\dagger_\downarrow)$ 
of field operators for up- and down-spin electrons. 
However, here we only consider the pairing between up- and down-spin electrons, 
neglecting spin-triplet equal spin pairing. 
Thus, the BCS Hamiltonian is reduced to $2 \times 2$ matrix, as  
\begin{eqnarray}
\int{\rm d}{\bf r} 
\left( \begin{array}{cc} \hat{\psi}_\uparrow^\dagger({\bf r}) &  \hat{\psi}_\downarrow({\bf r}) \\\end{array}\right) 
\left( \begin{array}{cc} 
K_\uparrow({\bf r}) & \Delta({\bf r}) \\ 
\Delta^\ast({\bf r}) & -K_\downarrow^\ast({\bf r}) \\  \end{array}\right) 
\left( \begin{array}{c} \hat{\psi}_\uparrow({\bf r}) \\  \hat{\psi}_\downarrow^\dagger({\bf r}) \\\end{array}\right) 
\nonumber \\ && 
\label{eq:Hamiltonian}
\end{eqnarray}
with 
$K_{\uparrow / \downarrow}({\bf r})
=(\hbar^2/2m^\ast) (-{\rm i}\nabla + \frac{\pi}{\phi_0}{\bf A} )^2
  \mp \mu_{\rm B}H -\mu$. 
We note that $K_\uparrow({\bf r}) \ne K_\downarrow({\bf r})$ 
in the presence of paramagnetic effect by the Zeeman energy. 
Following the method 
in Refs. \cite{Mizohata,Mizushima,Takahashi,IchiokaSDW,TakigawaSDW},  
by a unitary transformation 
\begin{eqnarray}
\left( \begin{array}{c} \hat{\psi}_\uparrow({\bf r}) \\  \hat{\psi}_\downarrow^\dagger({\bf r}) \\\end{array}\right) 
=\sum_{\varepsilon}
\left( \begin{array}{cc}
  u_{1 \varepsilon}({\bf r}) &  -v^\ast_{2 \varepsilon}({\bf r}) \\
  v_{1 \varepsilon}({\bf r}) &   u^\ast_{2 \varepsilon}({\bf r}) \\ \end{array}\right) 
\left( \begin{array}{c} \hat{\gamma}_{1 \varepsilon} \\  \hat{\gamma}_{2 \varepsilon} \\\end{array}\right),  
\end{eqnarray}
we diagonalize the Hamiltonian in Eq. (\ref{eq:Hamiltonian}) as 
\begin{eqnarray} && 
\left( \begin{array}{cc} 
  u_{1 \varepsilon'} &  -v^\ast_{2 \varepsilon'} \\
  v_{1 \varepsilon'} &   u^\ast_{2 \varepsilon'} \\ \end{array}\right)^{-1} 
\left( \begin{array}{cc} 
K_\uparrow({\bf r}) & \Delta({\bf r}) \\ 
\Delta^\ast({\bf r}) & -K_\downarrow^\ast({\bf r}) \\  \end{array}\right) 
\left( \begin{array}{cc}
  u_{1 \varepsilon} &  -v^\ast_{2 \varepsilon} \\
  v_{1 \varepsilon} &   u^\ast_{2 \varepsilon} \\ \end{array}\right) 
\nonumber \\ && 
= 
\left( \begin{array}{cc}
  E_{1 \varepsilon} &  0  \\
  0 &  -E_{2 \varepsilon} \\ \end{array}\right) \delta_{\varepsilon',\varepsilon}.
\label{eq:diagonal}
\end{eqnarray}
Label of the eigenstate, $\varepsilon$, is given by 
$\varepsilon=(k_x, k_y,i_z)$ in this work. 
From Eq. (\ref{eq:diagonal}), BdG equation is obtained as 
\begin{eqnarray}
\left( \begin{array}{cc}
K_\uparrow({\bf r}) & \Delta({\bf r}) \\ 
\Delta^\ast({\bf r}) & -K_\downarrow^\ast({\bf r}) \\  \end{array}\right) 
\left( \begin{array}{c}
  u_{ \epsilon}({\bf r}) \\   v_{\epsilon}({\bf r}) \\  \end{array}\right) 
=E_\epsilon 
\left( \begin{array}{c}
  u_{ \epsilon}({\bf r}) \\   v_{\epsilon}({\bf r}) \\  \end{array}\right) ,  
\end{eqnarray}
where 
\begin{eqnarray}
\left( \begin{array}{c}
  u_{ \epsilon}({\bf r}) \\   v_{\epsilon}({\bf r}) \\  \end{array}\right) 
=\left(\begin{array}{c}
  u_{1 \varepsilon}({\bf r}) \\   v_{1 \varepsilon}({\bf r}) \\  \end{array}\right) , 
 \left(\begin{array}{c}
  -v^\ast_{2 \varepsilon}({\bf r}) \\   u^\ast_{2 \varepsilon}({\bf r}) \\  \end{array}\right) 
\label{eq:unified}
\end{eqnarray}
for $E_\epsilon=E_{1 \varepsilon}$ and $-E_{2 \varepsilon}$, respectively. 
In the case of $K_\uparrow({\bf r})=K_\downarrow({\bf r})$, there are relations 
$E_{1 \varepsilon}=E_{2 \varepsilon}$, 
$u_{1 \varepsilon}({\bf r})=u_{2 \varepsilon}({\bf r})$, and 
$v_{1 \varepsilon}({\bf r})=v_{2 \varepsilon}({\bf r})$.  
In many cases solving the BdG equation, 
eigen states are divided to two groups: 
positive eigen energies $E_{1 \varepsilon}$ 
with ($u_{1 \varepsilon}({\bf r})$, $v_{1 \varepsilon}({\bf r})$),  
and negative ones $-E_{2 \varepsilon}$ 
with ($-v^\ast_{2 \varepsilon}({\bf r})$, $u^\ast_{2 \varepsilon}({\bf r})$).  
However, here we combine two groups of the eigen states 
by unified notation $E_{\epsilon}$, $u_{\epsilon}({\bf r})$, $v_{\epsilon}({\bf r})$ in Eq. (\ref{eq:unified}). 
There, eigen-states are labeled by $\epsilon=(\varepsilon,\alpha)$ with 
$\alpha$ for two states of $E_{1 \varepsilon}$ and $-E_{2 \varepsilon}$. 

By the unitary transformation, density of up- and down-spin electrons are calculated as 
\begin{eqnarray} && 
n_\uparrow({\bf r}) 
=\langle \hat\psi_\uparrow^\dagger({\bf r}) \hat\psi_\uparrow({\bf r}) \rangle 
\nonumber \\ && 
=\sum_\varepsilon \left\{ |u_{1 \varepsilon}({\bf r})|^2 f(E_{1 \varepsilon})
   +|v_{2 \varepsilon}^\ast({\bf r})|^2 f(-E_{2 \varepsilon}) \right\}, \qquad
\label{eq:n-up}
\\ && 
n_\downarrow({\bf r}) 
=\langle \hat\psi_\downarrow^\dagger({\bf r}) \hat\psi_\downarrow({\bf r}) \rangle 
\nonumber \\ && 
=\sum_\varepsilon \left\{ |v_{1 \varepsilon}({\bf r})|^2 f(-E_{1 \varepsilon})
   +|u_{2 \varepsilon}^\ast({\bf r})|^2 f(E_{2 \varepsilon}) \right\}, \qquad 
\label{eq:n-down}
\end{eqnarray}
where $\langle \cdots \rangle$ indicates the statistical average.
Similarly, $y$-component of up- and down-spin current is given by 
\begin{eqnarray} && 
J_\uparrow({\bf r}) 
={\rm Re}\langle \hat\psi_\uparrow^\dagger({\bf r})
                 \hat{F} \hat\psi_\uparrow({\bf r}) \rangle 
\nonumber \\ && 
=\sum_\varepsilon {\rm Re}\{ u^\ast_{1 \varepsilon}({\bf r})
                 \hat{F} u_{1 \varepsilon}({\bf r}) f(E_{1 \varepsilon})
\nonumber \\ && \qquad \qquad 
  -v_{2 \varepsilon}({\bf r}) \hat{F} 
   v^\ast_{2 \varepsilon}({\bf r}) f(-E_{2 \varepsilon}) \}, 
\label{eq:j-up}
\\ && 
J_\downarrow({\bf r}) 
={\rm Re}\langle \hat\psi_\downarrow^\dagger({\bf r})
          \hat{F} \hat\psi_\downarrow({\bf r}) \rangle 
\nonumber \\ && 
=\sum_\varepsilon {\rm Re}\{ v_{1 \varepsilon}({\bf r})
          \hat{F} v^\ast_{1 \varepsilon}({\bf r})  f(-E_{1 \varepsilon})
\nonumber \\ && \qquad \qquad 
         +u_{2 \varepsilon}^\ast({\bf r}) \hat{F} 
          u_{2 \varepsilon}({\bf r}) f(E_{2 \varepsilon}) \}, 
\label{eq:j-down}
\end{eqnarray}
with 
$\hat{F}=(e \hbar/m^\ast) (-{\rm i} \partial_y +\frac{\pi}{\phi_0}A_y )$.  
The pair potential is given by the gap equation 
\begin{eqnarray} && 
\Delta({\bf r}) \equiv 
V_{\rm pair}\langle \hat\psi_\uparrow({\bf r}) \hat\psi_\downarrow({\bf r})\rangle 
\nonumber \\ && 
=V_{\rm pair}{\sum_\varepsilon}'\left\{ 
  u_{1 \varepsilon}({\bf r}) v^\ast_{1 \varepsilon}({\bf r}) f(-E_{1 \varepsilon})
 -v^\ast_{2 \varepsilon}({\bf r})u_{2 \varepsilon}({\bf r}) f(E_{2 \varepsilon}) 
\right\}. 
\nonumber \\ && 
\label{eq:delta-updown}
\end{eqnarray}

Using the notation in Eq. (\ref{eq:unified}), 
Eqs. (\ref{eq:n-up})-(\ref{eq:delta-updown}) 
are written as 
\begin{eqnarray} && 
n_\uparrow({\bf r}) 
=\sum_\epsilon  |u_{\epsilon}({\bf r})|^2 f(E_{\epsilon}), 
\\ && 
n_\downarrow({\bf r}) 
=\sum_\epsilon  |v_{\epsilon}({\bf r})|^2 f(-E_{\epsilon}), 
\\ && 
\Delta({\bf r}) 
=V_{\rm pair}{\sum_\epsilon}'
 u_{\epsilon}({\bf r}) v^\ast_{\epsilon}({\bf r}) f(-E_{\epsilon}), 
\\ && 
J_\uparrow({\bf r}) 
=\sum_\epsilon {\rm Re}\left\{
  u^\ast_{\epsilon}({\bf r}) \hat{F} u_{\epsilon}({\bf r}) f(E_{\epsilon})
                      \right\}, \qquad
\\ && 
J_\downarrow({\bf r}) 
=\sum_\epsilon {\rm Re}\left\{
  v_{\epsilon}({\bf r}) \hat{F} v^\ast_{\epsilon}({\bf r})  f(-E_{\epsilon})
                      \right\}. \qquad 
\end{eqnarray}
Substituting Eq. (\ref{eq:uv-funs}) to these equations, 
we obtain Eqs. (\ref{eq:Delta1})-(\ref{nz}). 
This formulation of the BdG equation with 
$K_\uparrow({\bf r}) \ne K_\downarrow({\bf r})$ 
was used in previous studies for 
inhomogeneous superconducting state coexisting 
with the spin density wave~\cite{IchiokaSDW,TakigawaSDW}, 
and superfluid phase of trapped Fermion gas 
with population imbalance of up- and down-spins~\cite{Mizushima,Takahashi}.

 
\end{document}